\documentclass[16 pt]{article}
\usepackage{makeidx}
\usepackage{epsf}
\usepackage{times}
\usepackage{amsmath}

\begin{document}

\title{\bf
 {\em UDC 536.46 }\\ \vspace{1cm}\Large \bf Random Noise and
Pole-Dynamics \\in Unstable Front Propagation II}
\author{Oleg Kupervasser, Zeev Olami\\
Department of Chemical Physics\\The Weizmann Institute of
Science\\Rehovot 76100, Israel }

\maketitle

\begin{abstract}
\noindent
 The current paper is a corrected version of our previous
paper (Olami et al., PRE {\bf 55} (3),(1997)). Similarly to
previous version we investigate the problem of flame propagation.
This problem is studied as an example of unstable fronts that
wrinkle on many scales. The analytic tool of pole expansion in the
complex plane is employed to address the interaction of the
unstable growth process with random initial conditions and
perturbations. We argue that the effect of random noise is immense
and that it can never be neglected in sufficiently large systems.
We present simulations that lead to scaling laws for the velocity
and acceleration of the front as a function of the system size and
the level of noise, and analytic arguments that explain these
results in terms of the noisy pole dynamics.This version corrects
some very critical errors made in (Olami et al., PRE {\bf 55}
(3),(1997)) and makes more detailed description of excess number
of poles in system , number of poles that appear in the system in
unit of time, life time of pole. It allows us to understand more
correctly dependence of the system parameters on noise than in
(Olami et al., PRE {\bf 55} (3),(1997))

\end{abstract}


\section{Introduction}

This article considers the very interesting problem of describing
the nonlinear stage of development of hydrodynamic instability of
the flame. This problem can be considered for 1D (channel
propagation) , 2D (cylindrical case) , 3D (spherical case). The
direct numerical simulations can be made based on of the
Navier–Stokes equations including chemical kinetics in the form of
the Arrhenius law \cite{Liber}. For 2D and 3D cases we can see
experimentally observable effects \cite{89GIS}, \cite{11GIS} of
the self-acceleration of the front of a divergent flame, the
formation of cellular structure, and other effects.

Much more simple equation for flame front propagation can be
obtained \cite{77Siv}, \cite{94FSF} . It is Michelson-Sivashinsky
approximation model. The Michelson-Sivashinsky model assumes very
serious limitations, such as the smallness of the coefficient of
gas expansion and, consequently, the potential flow in the
combustion products and fresh mixture, weak nonlinearity, the
assumptions of the stabilizing effect of the curvature of the
flame front and a linear dependence on the curvature of the normal
speed, and others.  The calculations with a noise term was
performed in both one-dimensional and   two-dimensional
formulations of the problem in our previous papers \cite{prev},
 \cite{96KOP}, \cite{KOP951}, \cite{Kuper}, \cite{Kupopen}, \cite{prev1}
and recently by Karlin and Sivashinsky \cite{Kar1} , \cite{Kar2}
for 1D, 2D and 3D cases.

Interest of this model stems, firstly, from the fact that despite
the serious limitations, this model can qualitatively describe the
scenario of hydrodynamic instability and, in particular, the
self-acceleration of the front of a divergent flame, the formation
of a cellular structure, and other effects. Secondary, this
nonlinear model has exact solutions which can be constructed on
the basis of pole expansions \cite{82LC}, \cite{85TFH},
\cite{89Jou}, \cite{90Jou}. This pole expansion method got
development in the following papers. There was investigated
relationships between pole solutions and partial decomposition in
Fourier series \cite{Minaev1}, between the gas flow field and pole
solutions \cite{Minaev2}. The future development can be found in
our previous papers and the correspondent references inside of
\cite{prev}, \cite{96KOP}, \cite{KOP951}, \cite{Kuper},
\cite{Kupopen}, \cite{prev1}.

It must be mentioned that the simplest 1D case of the flame front
propagation is very important. It was the main reason for creation
this papers  considering in detail the 1D case. Such investigation
of this case allows us to understand qualitatively and
quantitatively the pole dynamics. This understanding is a basis
for the consideration more complex  2D and 3D cases. The our paper
\cite{KOP951} clearly demonstrate this fact. The  cellular
structure, acceleration exponent for 2D case was found on the
basis of 1D results.

The future development of our results is gotten in papers
\cite{Minaev3}, \cite{Minaev4}, \cite{Minaev5}. In the very
interesting paper \cite{Minaev3} the noise term was considered in
the poles-like form. The numerical results for 1D case are in a
good consistence with the theoretical results of this paper.

This paper is update version of our previous version \cite{prev}.
In this version we correct some error. For example, we choose
incorrect spectrum width of white noise with constant amplitude
during changing size of channel $L$. As result we didn't obtain in
\cite{prev} saturation of the frame front velocity during
increasing $L$ in contradiction with this paper. The second
important error connected to number of regimes defined on the
``phase diagram" of the system as a function of $L$ and $f$.
Indeed not three but four such regimes exist. The noise influence
theory developed in \cite{prev} described transition from regime I
to regime II and describe only these two regimes. The regime II
can not be observed because of numerical noise. The observable
regime III can not be explained by the help theory developed in
\cite{prev}. To make such explanation we calculate numerically and
try explain analytically such values as excess number of poles in
system , number of poles that appear in the system in unit of
time, life time of pole \cite{prev1}. The developed theory can
explain the existence of small dependence parameters of problem on
the noise in regime III.

The rest of the paper is organized as follows. We begin by
presenting equations of motion and pole-decomposition in the
channel geometry(Section~\ref{sec:method0}).Next
Section~\ref{sec:method1} describes acceleration of the flame
front, pole dynamics and noise. And finally
(Section~\ref{sec:method2}) we give summary and conclusions.

\section{Equations of Motion and Pole-decomposition in the Channel Geometry}\label{sec:method0}
It is known that planar flames freely propagating through
initially motionless homogeneous combustible mixtures are
intrinsically unstable. It was reported that such flames develop
characteristic structures which include cusps, and that under
usual experimental conditions the flame front accelerates as time
goes on. A model in $1+1$ dimensions that pertains to the
propagation of flame fronts in channels of width $\tilde L$ was
proposed in \cite{77Siv}. It is written in terms of position
$h(x,t)$ of the flame front above the $x$-axis. After appropriate
rescalings it takes the form:
\begin{equation}
{\partial h(x,t) \over \partial t}=
{1\over 2}\left[{\partial h(x,t) \over \partial x }\right]^2
 +\nu{\partial^2 h(x,t)\over \partial x^2}+ I\{h(x,t)\}+1 \ . \label{Eqnondim}
\end{equation}
The domain is  $0 <x< \tilde L$, $\nu$ is a parameter and we use periodic boundary
conditions. The functional $I[h(x,t)]$
is the Hilbert transform of derivative
which is conveniently defined in terms of the spatial
Fourier transform
\begin{eqnarray}
&&h(x,t)= \int_{-\infty}^{\infty} e^{i k x}\hat h(k,t) dk \label{Four}\\
&& I[h(k,t)] = |k| \hat h(k,t) \label{hil}
\end{eqnarray}
For the purpose of introducing the pole-decomposition it is convenient to
rescale the domain to $0<\theta <2\pi$. Performing this rescaling and
denoting the
resulting quantities with the same notation we have
\begin{eqnarray}
&&{\partial h(\theta,t) \over \partial t}=
{1\over 2L^2}\left[{\partial h(\theta,t) \over \partial \theta }\right]^2
 +{\nu\over L^2}{\partial^2 h(\theta,t)\over \partial\theta^2}\nonumber \\&&+
{1\over L}I\{h(\theta,t)\}+1 \ .
\label{Eqdim}
\end{eqnarray}
In this equation $L=\tilde L/2\pi$.
Next we change variables to $u(\theta,t)\equiv {\partial
h(\theta,t)/\partial\theta}$.
 We find
\begin{equation}
{\partial u(\theta,t) \over \partial t}=
{u(\theta,t)\over L^2}{\partial u(\theta,t) \over \partial \theta }
 +{\nu\over L^2}{\partial^2 u(\theta,t)\over \partial \theta^2}+ {1\over
L}I\{u(\theta,t)\} \ . \label{eqfinal}
\end{equation}
It is well known that the flat front solution of this equation is linearly
unstable.
The linear spectrum in $k$-representation is
\begin{equation}
\omega_k=|k|/L-\nu k^2/L^2 \ . \label{spec}
\end{equation}
There exists a
typical scale $k_{max}$ which is the last unstable mode
\begin{equation}
k_{max} = {L\over \nu} \ . \label{kmax}
\end{equation}
Nonlinear effects stabilize a new steady-state which is discussed next.

The outstanding feature of the solutions of this equation is the
appearance of cusp-like structures in the developing fronts.
Therefore a representation in terms of Fourier modes is very
inefficient. Rather, it appears very worthwhile to represent such
solutions in terms of sums of functions of poles in the complex
plane. It will be shown below that the position of the cusp along
the front is determined by the real coordinate of the pole,
whereas the height of the cusp is in correspondence with the
imaginary coordinate. Moreover, it will be seen that the dynamics
of the developing front can be usefully described in terms of the
dynamics of the poles. Following \cite{96KOP,82LC,85TFH,90Jou} we
expand the solutions $u(\theta,t)$ in functions that depend on $N$
poles whose position $z_j(t)\equiv x_j(t)+iy_j(t)$ in the complex
plane is time dependent:
\begin{eqnarray}
&&u(\theta,t)=\nu\sum_{j=1}^{N}\cot \left[{\theta-z_j(t) \over 2}\right]
   + c.c.\nonumber \\
&&=\nu\sum_{j=1}^{N}{2\sin [\theta-x_j(t)]\over
\cosh [y_j(t)]-\cos [\theta-x_j(t)]}\ , \label{upoles}
\end{eqnarray}
\begin{equation}
h(\theta,t)=2\nu\sum_{j=1}^{N}{\ln \Big[\cosh (y_j(t))-\cos (\theta-x_j(t))
\Big]}+C(t) \ . \label{rpoles}
\end{equation}
In (\ref{rpoles}) $C(t)$ is a function of time. The function (\ref{rpoles})
is a superposition of quasi-cusps (i.e. cusps that are rounded at the tip). The
real part of the pole position (i.e. $x_j$) is the coordinate (in the domain
$[0,2\pi]$) of the maximum
of the quasi-cusp, and the imaginary part of the pole position (i.e $y_j$)
is related to
the depth of the quasi-cusp. As $y_j$ decreases the depth of the cusp
increases. As $y_j \to 0$ the depth diverges to infinity. Conversely, when $y_j\to
\infty$ the
depth decreases to zero.

The main advantage of this representation is that the propagation and
wrinkling of the
front can be described via the dynamics of the poles. Substituting
(\ref{upoles}) in (\ref{eqfinal}) we derive the following ordinary
differential equations for the positions of the poles:
\begin{equation}
- L^2{dz_{j}\over dt}=\Big[\nu\sum_{k=1
  ,k\neq j}^{2N }\cot \left({z_j-z_k\over 2}\right)
  +i{L\over 2 }sign [Im(z_j)]\Big].\label{eqz}
\end{equation}
We note that in (\ref{upoles}), due to the complex conjugation, we have $2N$ poles
which are arranged in pairs such that for $j<N$ $z_{j+N}=\bar z_j$. In the
second sum in (\ref{upoles}) each pair of poles contributed one term. In Eq.(\ref{eqz})
we again employ $2N$ poles since all of them interact. We can write the
pole dynamics in  terms of the real and imaginary parts $x_j$ and $y_j$.
Because of the
arrangement in pairs it is sufficient to write the equation for either
$y_j>0$ or for
$y_j<0$. We opt for the first. The equations for the positions of the poles read
\begin{eqnarray}
&&-L^2{dx_{j}\over dt}=\nu\sum_{k=1,k\neq j}^N
   \sin(x_j-x_k)\Bigg[
   [\cosh (y_j-y_k) \label{xj} \\ && -\cos (x_j-x_k)]^{-1}+[\cosh
(y_j+y_k)-\cos (x_j-x_k)]^{-1}\Bigg]  \nonumber\\
&& L^2{dy_{j}\over dt}=\nu\sum_{k=1,k\neq j}^{N }\Big({\sinh (y_j-y_k)\over
   \cosh (y_j-y_k)-\cos (x_j-x_k)}\nonumber \\ &&+
   {\sinh (y_j+y_k)\over \cosh (y_j+y_k)-\cos (x_j-x_k)}
   \Big)+\nu\coth (y_j)- L\label{yj}  .
\end{eqnarray}
We note that if the initial conditions of the differential
equation (\ref{eqfinal}) are expandable in a finite number of
poles, these equations of motion preserve this number as a
function of time. On the other hand, this may be an unstable
situation for the partial differential equation, and noise can
change the number of poles. This issue will be examined at length
in Section \ref{sec:method1}. We turn now to a discussion of the
steady state solution of the equations of the pole-dynamics.
\subsection{Qualitative properties of the stationary solution}\label{sec:met0}

The steady-state solution of the flame front propagating in channels of
width $2\pi$
was presented in Ref.\cite{85TFH}. Using these results we can immediately
translate
the discussion to a channel of width $L$. The main results are summarized
as follows:
\begin{enumerate}
\item There is only one
stable stationary solution which is geometrically represented by a giant
cusp (or equivalently one finger) and
analytically by $N(L)$ poles which are aligned on one line parallel to the
imaginary
axis. The existence of this solution is made clearer with the following remarks.
\item There exists an attraction between the poles
along the real line. This is obvious from Eq.(\ref{xj}) in which the sign
of $dx_j/dt$ is
always determined by $\sin(x_j-x_k)$. The resulting dynamics merges all the $x$
positions of poles whose $y$-position remains finite.
\item The $y$
positions are distinct, and the poles are aligned above each others in positions
$y_{j-1}<y_j<y_{j+1}$ with the maximal being $y_{N(L)}$. This can be understood
from Eq.(\ref{yj}) in which the interaction is seen to be repulsive at
short ranges,
but changes sign at longer ranges.
\item If one adds
an additional pole to such
a solution, this pole (or another) will be pushed to infinity along the
imaginary axis.
If the system has less than $N(L)$ poles it is unstable to the addition of
poles,
and any noise will drive the system towards this unique state. The number
$N(L)$ is
\begin{equation}
N(L)= \Big[{1 \over 2}\left( {L \over \nu }+1\right) \Big]\ , \label{NofL}
\end{equation}
where $\Big[ \dots \Big]$ is the integer part. To see this consider
a system with $N$ poles and such
that all the values of $y_j$ satisfy the condition $0< y_j <y_{max}$.
Add now one additional pole whose coordinates are $z_a\equiv (x_a,y_a)$ with
$y_a\gg
y_{max}$. From the equation of motion for $y_a$, (\ref{yj}) we see that
the terms in the sum are all of the order of unity as is also the
$\cot(y_a)$ term. Thus
the equation of motion of $y_a$ is approximately
\begin{equation}
{dy_a \over dt}\approx \nu{2N+1 \over L^2}-{1\over L} \ . \label{ya}
\end{equation}
The fate of this pole depends on the number of other poles.
If $N$ is too large the pole will run to infinity, whereas if $N$ is  small
the pole
will be attracted towards the real axis. The condition for moving away to infinity
is that $N > N(L)$ where $N(L)$ is given by (\ref{NofL}). On the
other hand the $y$ coordinate of the poles cannot hit zero.
Zero is a repulsive line, and poles are pushed away from zero with infinite velocity.
 To see this consider
a pole whose $y_j$ approaches zero. For any finite $L$ the term
$\coth(y_j)$ grows
unboundedly whereas all the other terms in Eq.(\ref{yj}) remain  bounded.
\item  The height of the cusp is proportional to $L$. The distribution of
positions
of the poles along the line of constant $x$ was worked out in \cite{85TFH}.
\end{enumerate}
We will refer to the solution with all these properties as the
Thual-Frisch-Henon
(TFH)-cusp solution.

\subsection{Nonlinear Stability}\label{sec:met1}

The intuition gained so far can be used to discuss the issue of
stability of a stable system to {\em larger} perturbations. In
other words, we may want to add to the system poles at finite
values of $y$ and ask about their fate. We first show in this
subsection that poles whose initial $y$ value is below
$y_{max}\sim \log (L^2/\nu^2)$ will be attracted towards the real
axis. The scenario is similar to the one described in the last
paragraph.

Suppose that we generate a stable system with a giant cusp at
$\theta_c=0$ with poles distributed along the $y$ axis up to
$y_{max}$. We know that the sum of all the forces that act on the
upper  pole is zero. Consider then an additional pole inserted in
the position $(\pi,y_{max})$. It is obvious from Eq.(\ref{yj})
that the forces acting on this pole will pull it downward. On the
other hand if its initial position is much above $y_{max}$ the
force on it will be repulsive towards infinity. We see that this
simple argument identifies $y_{max}$ as the typical scale for
nonlinear instability.

Next we estimate $y_{max}$ and interpret our result in terms of
the {\em amplitude} of a perturbation of the flame front. We
explained that uppermost pole's position fluctuates between a
minimal value and infinity as $L$ is changing. We want to estimate
the characteristic scale of the minimal value of $y_{max}(L)$. To
this aim we employ the result of ref.\cite{85TFH} regarding the
stable distribution of pole positions in a stable large system.
The parametrization of \cite{85TFH} differs from ours; to go from
our parametrization in Eq.(\ref{eqfinal}) to theirs we need to
rescale $u$ by $L^{-1}$ and $t$ by $L$. The parameter $\nu$ in
their parameterization is $\nu/L$ in ours. According to
\cite{85TFH} the number of poles between $y$ and $y+dy$ is given
by the $\rho(y)dy$ where the density $\rho(y)$ is
\begin{equation}
\rho(y)={L\over \pi^2\nu}\ln[\coth(|y|/4)] \ . \label{dist}
\end{equation}
To estimate the minimal value of $y_{max}$ we require that the
tail of the distribution $\rho(y)$ integrated between this value
and infinity will allow one single pole. In other words,
\begin{equation}
\int_{y_{max}}^\infty dy \rho(y) \approx 1 \ . \label{integ}
\end{equation}
Expanding (\ref{dist}) for large $y$ and integrating explicitly
the result in (\ref{integ}) we end up with the estimate
\begin{equation}
y_{max} \approx 2\ln\Big[{4L\over \pi^2\nu}\Big]
\end{equation}
For large $L$ this result is $y_{max}\approx \ln({L^2 \over
\nu^2})$. If we now add an additional pole in the position
$(\theta,y_{max})$ this is equivalent to perturbing the solution
$u(\theta,t)$ with a function $\nu e^{-y_{max}} \sin(\theta)$, as
can be seen directly from (\ref{upoles}). We thus conclude that
the system is unstable to a perturbation {\em larger} than
\begin{equation}
u(\theta) \sim \nu^3 \sin(\theta)/L^2 \ . \label{nu3L2}
\end{equation}
This indicates a very strong size dependence of the sensitivity of
the giant cusp solution to external perturbations. This will be an
important ingredient in our discussion of noisy systems.

\section{Acceleration of the Flame Front, Pole Dynamics and Noise}\label{sec:method1}
A major motivation of this Section is the observation that in
radial geometry the same equation of motion shows an acceleration
of the flame front. The aim of this section is to argue that this
phenomenon is caused by the noisy generation of new poles.
Moreover, it is our contention that a great deal can be learned
about the acceleration in radial geometry by considering the
effect of noise in channel growth. In Ref. \cite{85TFH} it was
shown that any initial condition which is represented in poles
goes to a unique stationary state which is the giant cusp which
propagates with a constant velocity $v=1/2$ up to small $1/L$
corrections. In light of our discussion of the last section we
expect that any smooth enough initial condition will go to the
same stationary state. Thus if there is no noise in the dynamics
of a finite channel, no acceleration of the flame front is
possible. What happens if we add noise to the system?

For concreteness we introduce an additive white-noise term $\eta(\theta,t)$to
the equation of motion  (\ref{eqfinal})  where
\begin{equation}
\eta(\theta,t) = \sum_k{\eta_k(t) \exp{(ik\theta)}}\ , \label{eta}
\end{equation}
and the Fourier amplitudes $\eta_k$ are correlated according to
\begin{equation}
<\eta_k(t)\eta^{*}_{k'}(t')>={f \over L}
\delta_{k,k'}\delta(t-t') \ . \label{corr}
\end{equation}
We will first examine the result of numerical simulations of noise-driven
dynamics,
and later return to the theoretical analysis.
\subsection{Noisy Simulations}



\begin{figure}
\epsfxsize=9.0truecm \epsfbox{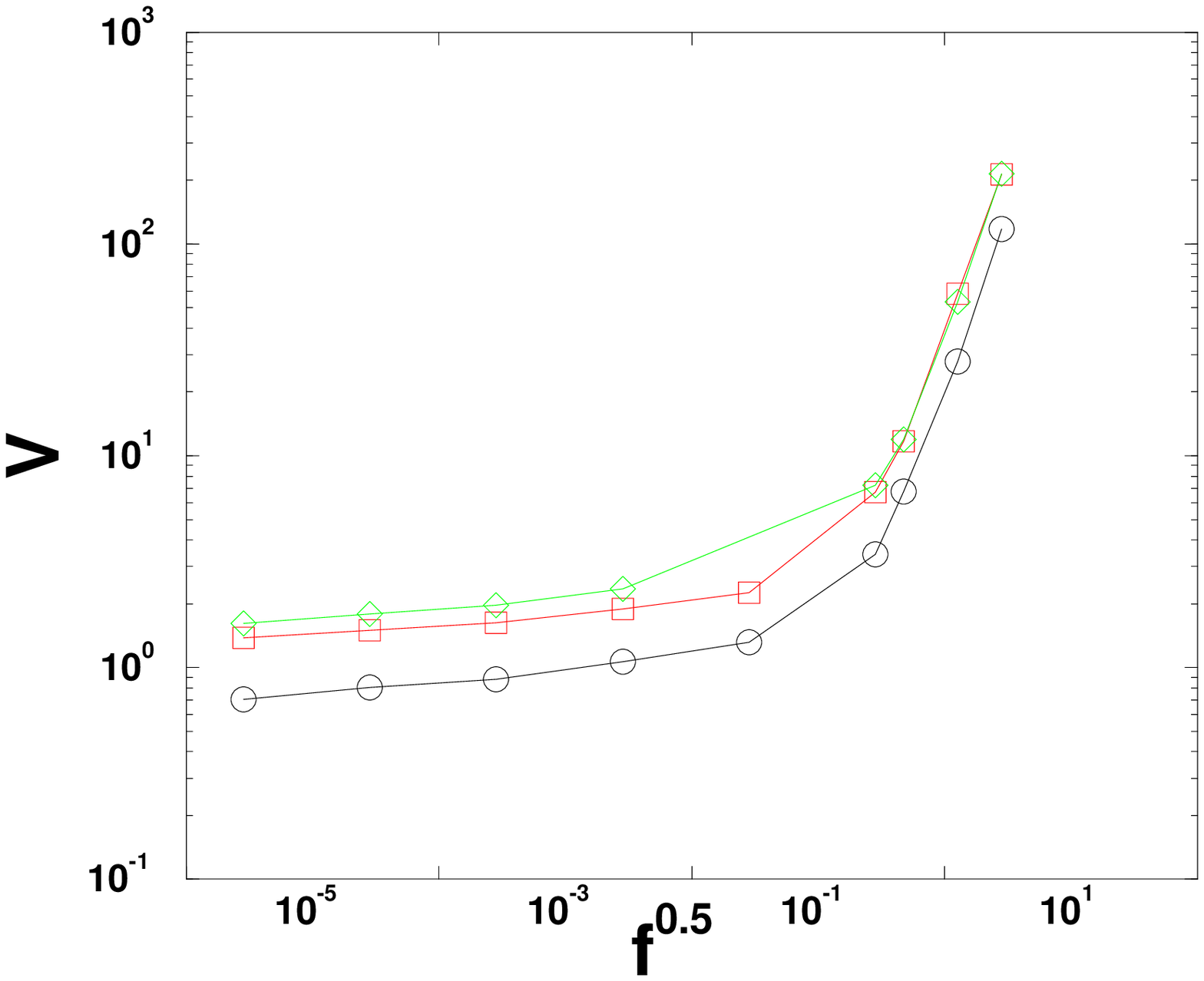} \caption {}
\label{file=Fig.2}
\end{figure}

Previous numerical  investigations \cite{94FSF,90GS} did not introduce noise in
a controlled
fashion. We will argue later that some of the phenomena encountered in
these simulations
can be ascribed to the (uncontrolled) numerical noise. We performed numerical
simulations of Eq.(\ref{eqfinal} using a pseudo-spectral method. The time-stepping scheme
was chosen as Adams-Bashforth with 2nd order presicion in time. The additive white noise was
generated in Fourier-space by choosing
$\eta_k$
for every $k$ from a flat distribution
in the interval $[-\sqrt{2{f \over L}},\sqrt{2{f \over L}}]$.
We examined the average steady state
velocity of the front
as a function of $L$ for fixed $f$ and as a function of $f$ for fixed $L$.
We found the
interesting phenomena that are summarized here:
\begin{enumerate}
\item In Fig.\ref{file=Fig.2} we can see two different regimes of
behavior the average velocity $v$ as function of noise $f^{0.5}$
for fixed system size L. For the noise $f$ smaller then same fixed
value $f_{cr}$
\begin{equation}
v\sim f^\xi \ . \label{vf}
\end{equation}
For these values of $f$ this dependence is very weak, and
$\xi\approx 0.02$.
 For
 large values of $f$ the dependence is much stronger
\item In Fig.\ref{file=Fig.1} we can see growth of the average
velocity $v$ as function of the system size L. After some values
of L we can see saturation of the velocity. For regime $f<f_{cr}$
the growth of the velocity can be written as
\begin{equation}
v\sim L^\mu , \quad \mu\approx 0.40\pm 0.05 \ . \label{scale1}
\end{equation}

\end{enumerate}
\subsection{Calculation  of Poles Number in the System}
The interesting problem that we would like to solve here it is to
find number of poles that exist in our system outside the giant
cusp. We can make it by next way: to calculate number of cusps
(points of minimum or inflexional points) and their position on
the interval $ \theta:[0,2\pi]$ in every moment of time and to
draw positions of cusp like function of time, see
Fig.\ref{file=Fig.3}.


We assume that our system is almost
all time in "quasi-stable" state, i.e. every new cusp that appears in
the system includes only one pole.
By help pictures obtained by such way we can find
\begin{enumerate}
\item By calculation number of cusp in some moment of time and
by investigation of history of every cusp (except the giant cusp)
, i.e. how many initial cusps
take part in formation this cusp, after averaging with respect to
different moments of time we can find  mean number of poles that
exist  in our system outside the giant cusp.
Let us denote this number $\delta N$.
We can see four regimes that can be define with respect to dependence of
this number on noise $f$:

(i) Regime I: Such small noise that
no poles exist in our system outside the giant cusp.

(ii)
Regime II Strong dependence of poles number  $\delta N$  on noise $f$.

(iii) Regime III Saturation  poles number  $\delta N$  on noise $f$,
so we see very small dependence of this number on noise

\begin{equation}
\delta N \sim   f^{0.03}
\end{equation}

Saturated value of $\delta N$ is defined by next formula

\begin{equation}
\delta N \approx   N(L)/2 \approx {1 \over 4} {L \over \nu}
\end{equation}

where $N(L) \approx {1 \over 2} {L \over \nu}$ is number of poles in giant
cusp.

(iv) Regime IV We again see strong dependence of poles number  $\delta N$
on noise $f$.

\begin{equation}
\delta N \sim   f^{0.1}
\end{equation}

Because of numerical noise we can see in most of simulations only regime III
and IV. In future if we don't note something different we discuss regime III.

\item By calculation of new cusp number we can find number of poles
that appear in the system in unit of time ${ dN \over dt}$.
In regime III

\begin{equation}
{ dN \over dt} \sim   f^{0.03}
\end{equation}

Dependence on $L$ and $\nu$ define by

\begin{equation}
{ dN \over dt} \sim  L^{0.8}
\end{equation}

\begin{equation}
{ dN \over dt} \sim  {1 \over \nu^2}
\end{equation}

And in regime IV

\begin{equation}
{ dN \over dt} \sim   f^{0.1}
\end{equation}
\end{enumerate}
\subsection{Theoretical Discussion of the Effect of Noise}
\subsubsection{The Threshold of Instability to Added Noise. Transition
from regime I to regime II}
\label {regime0}
First we present the theoretical arguments that explain the sensitivity of the
giant cusp solution to the effect of added noise. This sensitivity increases
dramatically with increasing the system size $L$. To see this we use again the
relationship between the linear stability analysis and the pole dynamics.

Our additive noise introduces perturbations with all $k$-vectors.
We showed previously that the most unstable mode is the $k=1$
component $A_1 \sin(\theta)$. Thus the most effective noisy
perturbation is $\eta_1 \sin(\theta)$ which can potentially lead
to a growth of the most unstable mode. Whether or not this mode
will grow depends on the amplitude of the noise. To see this
clearly we return to the pole description. For small values of the
amplitude $A_1$ we represent $A_1 \sin(\theta)$ as a single pole
solution of the functional form $\nu e^{-y}\sin{\theta}$. The $y$
position is determined from $y=-\log{|A_1| /\nu}$, and the
$\theta$-position is $\theta=\pi$ for positive $A_1$ and
$\theta=0$ for negative $A_1$. For very small $A_1$ the fate of
the pole is to be pushed to infinity, independently of its
$\theta$ position; the dynamics is symmetric in $A_1\to -A_1$ when
$y$ is large enough. On the other hand when the value of $A_1$
increases the symmetry is broken and the $\theta$ position and the
sign of $A_1$ become very important. If $A_1>0$ there is a
threshold value of $y$ below which the pole is attracted down. On
the other hand if $A_1<0$, and $\theta=0$ the repulsion from the
poles of the giant cusp grows with decreasing $y$. We thus
understand that qualitatively speaking the dynamics of $A_1$ is
characterized by an asymmetric ``potential" according to
\begin{eqnarray}
\dot A_1 &=& -{\partial V(A_1)\over \partial A_1}\ , \label{dvda}\\
V(A_1) &=& \lambda A_1^2 -aA_1^3+\dots \ . \label{poten}
\end{eqnarray}

From the linear stability analysis we know that $\lambda\approx
\nu/L^2$, cf. Eq.(\ref{ya}). We know further that the threshold
for nonlinear instability is at $A_1\approx \nu^3/L^2$, cf.
Eq(\ref{nu3L2}). This determines that value of the coefficient
$a\approx 2/3\nu^2$. The magnitude of the ``potential" at the
maximum is
\begin{equation}
V(A_{max}) \approx \nu^7/L^6 \ . \label{vmax}
\end{equation}
The effect of the noise on the development of the mode $A_1\sin{\theta}$ can be
understood from the following stochastic equation
\begin{equation}
\dot A_1 = -{\partial V(A_1)\over \partial A_1}+\eta_1(t) \ . \label{stochA}
\end{equation}
It is well known \cite{Ris} that for such dynamics the rate of escape $R$
over the ``potential" barrier
for small noise is proportional to
\begin{equation}
R\sim {\nu\over L^2} \exp^{-\nu^7/{f \over L}L^6} \ . \label{wow}
\end{equation}
The conclusion is that any arbitrarily tiny noise becomes effective when the
system size increase and when $\nu$ decreases. If we drive the system with noise
of amplitude ${f \over L}$ the system can always be sensitive to this noise when its size
exceeds a critical value $L_c$ that is determined by ${f \over L_c} \sim \nu^7/L_c^6$.
This formula defines transition from regime I (no poles) to regime II.
For $L>L_c$
the noise will introduce new poles into the system.
Even numerical noise in simulations involving large size systems may have
a macroscopic influence.

The appearance of new poles must increase the velocity of the
front. The velocity is proportional to the mean of $(u/L)^2$. New
poles distort the giant cusp by additional smaller cusps on the
wings of the giant cusp, increasing $u^2$. Upon increasing the
noise amplitude more and more smaller cusps appear in the front,
and inevitably the velocity increases. This phenomenon is
discussed quantitatively in Section \ref{sec:method1}.
\subsubsection{Verifying of asymmetric "potential" form}
From the equations of the motion for poles we can find the
distribution of poles in the giant cusp \cite{85TFH}. If we know
the distribution of poles in the giant cusp we can then find the
form of the "potential" and verify numerically expressions for
values $\lambda$, $A_{max}$ and ${\partial V(A_1) \over \partial
A_1}$  discussed previously. The connection between amplitude
$A_1$ and the position of the pole $y$ is defined by $A_1=4 \nu
e^{-y}$ and the connection between the potential function $
{\partial V(A_1) \over \partial A_1}$ and the position of the pole
$y$ is defined by formula ${\partial V(A_1) \over \partial A_1}= 4
\nu {dy \over dt} e^{-y}$, where ${dy \over dt}$ can be determined
from the equation of the motion of the poles. We can find
$A_{max}$ as the zero-point of ${\partial V(A_1) \over \partial
A_1}$ and $\lambda$ can be found as ${1 \over 2} {\partial^2
V(A_1) \over \partial A_1^2}$ for $A_1=0$. Numerical measurements
were made for the set of values $ L=2 n \nu$, where $n$ is a
integer and $ n>2$. For our numerical measurements we use  the
constant $\nu=0.005$ and the variable $L$, where $L$ changes in
the interval [1,150], or
 variable $\nu$ that changes in the interval [0.005,0.05] and the
 constant $L=1$.
The  results obtained follow:
\begin{enumerate}
\item
Formula for ${ A_{max} L^2 \over \nu^3}$
\begin{eqnarray}
{ A_{max} L^2 \over \nu^3}\approx 6.5 \ . \label{potenw1}
\end{eqnarray}
\item Formula for ${A_{max} \over A_{N(L)}}$
\begin{eqnarray}
{A_{max} \over A_{N(L)}} \approx 0.465 \ .\label{potenw2}
\end{eqnarray}
where $A_{N(L)}$ is defined by position of the upper pole.
\item Formula for ${\lambda L^2 \over \nu}$
\begin{eqnarray}
 {\lambda L^2 \over \nu}= 0.5 \ .\label{potenw3}
\end{eqnarray}
\item We also verify the boundary between regime I (no new cusps)
and regime II (new cusps appear). Fig. \ref{file=Fig.4} shows the
dependence of ${f \over L_c}$ on $L_c$. We can see that ${f/ L_c}
\sim 1/L_c^6$.

\end{enumerate}
These results are in good agreement with the theory.
\subsubsection{The Noisy Steady State and its Collapse with
Large Noise and System Size}
In this subsection we discuss the response of the giant cusp solution to
noise levels that are able to introduce a large number of excess poles in
addition
to those existing in the giant cusp. We will denote the excess number of poles
by $\delta N$. The first question that we address is how difficult is it to
insert yet an additional pole when there is already a given excess $\delta
N$. To
this aim we estimate the effective potential $V_{\delta N}(A_1)$ which is
similar to
(\ref{poten}) but is taking into account the existence of an excess number
of poles.
A basic approximation that we employ is that the fundamental form of the
giant cusp solution is not seriously modified by the existence of an excess
number
of poles. Of course this approximation breaks down quantitatively already
with one excess pole.
Qualitatively however it holds well until the excess number of
poles
is of the order of the original number $N(L)$ of the giant cusp solution.
Another approximation is that the rest of the linear modes play no role in this case.
At this
point we limit the discussion therefore to the situation $\delta N\ll N(L)$
(regime II).

To estimate the parameter $\lambda$ in the effective potential we consider the
dynamics of one pole whose $y$ position $y_a$ is far above $y_{max}$. According
to Eq.(\ref{ya}) the dynamics reads
\begin{equation}
{dy_a\over dt}\approx {2\nu (N(L)+\delta N)\over L^2} -{1\over L}
\end{equation}
Since the $N(L)$ term cancels against the $L^{-1}$ term (cf. Sec.
\ref{sec:met0}), we remain with a repulsive term that in the
effective potential translates to
\begin{equation}
\lambda={\nu\delta N\over L^2} \ . \label{lambda2}
\end{equation}
Next we estimate the value of the potential at the break-even
point between attraction and repulsion. In the last subsection we
saw that a foreign pole has to be inserted below $y_{max}$ in
order to be attracted towards the real axis. Now we need to push
the new pole below the position of the existing  pole whose index
is $N(L)-\delta N$. This position is estimated as in Sec
\ref{sec:met1} by employing the TFH distribution function
(\ref{dist}). We find
\begin{equation}
y_{\delta N}\approx 2\ln{\Big[{4L\over \pi^2\nu\delta N}\Big]} \ . \label{ydelN}
\end{equation}
As before, this implies a threshold value of the amplitude of single pole
solution
$A_{max}\sin{\theta}$ which is obtained from equating $A_{max}=\nu
e^{-y_{\delta N}}$.
We thus find in the present case $A_{max}\sim \nu^3(\delta N)^2/L^2$. Using
again
a cubic representation for the effective potential we find $a=2/(3\nu^2\delta
N)$ and
\begin{equation}
V(A_{max}) = {1\over 3}{\nu^7(\delta N)^5\over L^6}\ . \label{max2}
\end{equation}
Repeating the calculation of the escape rate over the potential barrier we find
in the present case
\begin{equation}
R\sim {\nu\delta N\over L^2} \exp^{-\nu^7(\delta N)^5/{f \over L}L^6} \ . \label{wow2}
\end{equation}

For a given noise amplitude ${f \over L}$ there is always a value of $L$ and $\nu$ for
which
the escape rate is of $O(1)$ as long as $\delta N$ is not too large. When
$\delta N$
increases the escape rate decreases, and eventually no additional poles can
creep
into the system. The typical number $\delta N$ for fixed values of the
parameters is
estimated from equating the argument in the exponent to unity:
\begin{equation}
\delta N\approx \left({f \over L}L^6/\nu^7\right)^{1/5} \ . \label{deltaN}
\end{equation}

We can see that $\delta N$ depend on noise $f$ very seriously.
It is not the case in regime III. Let us find conditions of transition
from regime II to regime III, where we see saturation of $\delta N$
with respect to noise $f$.

(i)  We use for the amplitude of pole solution that really
equal to ${2\nu \sin \theta \over \cosh(y_{\delta N}) - \cos \theta }$
expression $A_{max}=4 \nu  e^{-y_{
\delta N}}$ but it is right only for big number $y_{\delta N}$. For
$y_{\delta N} < 1 $ better approximation is $A_{max}=
{4 \nu \over y_{\delta N}^2}$.
From equation (\ref{ydelN}) we can find that the boundary value
$y_{\delta N}=1$ correspond to $\delta N \approx N(L)/2$

(ii) We use expression $y_{\delta N} \approx
2\ln{\Big[{4L\over \pi^2\nu\delta N}\Big]}$ but for big value of
$\delta N$ better approximation that can be find the same way is
$y_{\delta N} \approx {\pi^2 \nu \over 2 L}(N(L)-\delta N)
\ln {\Big[{ 8eL \over \pi^2 \nu (N(L)-\delta N)}\Big]}$.
These expressions give us  nearly equal result for $\delta N \approx N(L)/2$.

From (i) and (ii) we can make next conclusions

(a) Transition from regime II to regime III happens for nearly
$\delta N \approx N(L)/2$

(b) Using new expression in (i) and (ii) for amplitude $A_{max}$ and
$y_{\delta N}$ we can find for noise ${f \over L}$ in regime III:

\begin{equation}
{f \over L} \sim V(A_{max}) \sim \lambda A_{max}^2 \sim {\nu \delta N \over
L^2} ({4\nu \over y_{\delta N}^2})^2 \sim {L^2 \over \nu}  {\delta N
\over (N(L)-\delta N)^4}
\end{equation}

This expression define very slow dependence of $\delta N$ on noise ${f \over L}$
for $\delta N > N(L)/2$ that explain  noise saturation of $\delta N$
for regime III.

(c) Form of the giant cusp solution is defined by poles that are
closely to zero with respect to $y$.
For regime III $N(L)/2$ poles that have position
$y < y_{\delta N=N(L)/2}=1$ stay on these place. This result explain
why giant cusp solution is not seriously modified for regime III.

From Eq.(\ref{deltaN}) by help of boundary condition
\begin{equation}
\delta N \approx
N(L)/2 \label{polla}
\end{equation}
boundary noise $f_b$ between regime II and III can be found

\begin{equation}
f_b \sim { \nu^2 } \label{nose}
\end{equation}

The basic equation describing pole dynamics is next

\begin{equation}
{dN \over dt}={\delta N \over T} \label{tlif1}
\end{equation}

where ${dN \over dt}$ is number of poles that appear in unit of time
in our system, $\delta N$ is excess number of poles, T is mean life
time of pole
 (between appearing and merging with giant cusp). Using result of numerical
simulations for ${dN \over dt}$ and (\ref{polla}) we can find for $T$

\begin{equation}
T = {\delta N \over {dN \over dt}} \sim \nu L^{0.2} \label{tlif}
\end{equation}

So life time proportional to $\nu$ and depend on system size $L$
very weakly.

Moreover, the lifetime of a pole is defined by the lifetime of the
poles that are in a cusp. From the maximum point of the linear
part of Eq.(\ref{Eqnondim} ), we can find the mean character size

\begin{equation}
\lambda_m \sim \nu
\end{equation}

that defines the  size of our cusps. This result was confirmed in
numerical calculations executed in \cite{jula2}. Indeed, we can
see it from Fig.9 in \cite{jula2}. The mean number of poles in a
cusp

\begin{equation}
n_{big} \approx {\lambda_m \over 2 \nu} \sim const
\end{equation}

does not depend on $L$ and $\nu$. The mean  number of cusps is

\begin{equation}
N_{big} \sim {\delta N \over n_{big}} \sim {L \over \nu} \ .
\end{equation}

Let us assume that some  cusp exists in the main minimum of the
system.
 The lifetime of a pole in such a cusp is defined by three parts.

(I) Time of the cusp formation. This time is proportional to the
cusp size (with $\ln$-corrections) and the pole number in the cusp
(from pole motion equations)

\begin{equation}
T_1 \sim \lambda_m n_{big} \sim \nu
\end{equation}

(II) Time that the cusp is in the minimum neighborhood. This time
is defined by

\begin{equation}
T_2 \sim {a \over v} \,
\end{equation}

where $a$ is a  neighborhood of minimum, such that the force from
the giant cusp is smaller than the force from the fluctuations of
the excess pole number $\delta N$, and $v$ is the velocity of a
pole in this neighborhood. Fluctuations of excess pole number
$\delta N$ are expressed as

\begin{equation}
N_{fl}=\sqrt{\delta N} \ .
\end{equation}

From this result and the pole motion equations we find that

\begin{equation}
v \sim {\nu \over L} N_{fl} \sim {\nu \over L} \sqrt{{L \over
\nu}} \sim \sqrt{{\nu  \over L}} \ .
\end{equation}

The velocity from the giant cusp is defined by

\begin{equation}
v \sim {\nu \over L} N(L) {a \over L}  \sim {a \over L} \ .
\end{equation}

So from  equating these two equations we obtain

\begin{equation}
a \sim \sqrt{ \nu L}  \ .
\end{equation}

Thus for $T_2$ we obtain

\begin{equation}
T_2 \sim {a \over v} \sim L \ .
\end{equation}

(III) Time of attraction to the giant cusp. From the equations of
motion for the poles we get

\begin{equation}
T_3 \sim L \ln ({L \over a}) \sim L \ln \sqrt{L} \sim L \ .
\end{equation}

The investigated domain of the system size was found to be

\begin{equation}
T_1 \gg T_2,T_3
\end{equation}

Therefore full lifetime is

\begin{equation}
T=T_1 +T_2 +T_3 \sim \nu + s L \ ,
\end{equation}

where $s$ is a constant and

    \begin{equation}
    0 < s \ll 1 \ .
    \end{equation}

This result qualitatively and partly quantatively explains
dependence (\ref{tlif}). From (\ref{tlif}), (\ref{tlif1}),
(\ref{polla}) we can see that in regime III   ${dN \over dt}$ is
saturated with the system size $L$.

\subsection{The acceleration of the flame front due to noise}
\label{accel}
In this section we estimate the scaling exponents that characterize the velocity
of the flame front as a function of the system size. Our arguments in this
section
are even less solid than the previous ones, but nevertheless we believe that we succeed
to capture
some of the essential qualitative physics that underlies the interaction between
noise and instability and which results in the acceleration of the flame front.

To estimate the velocity of the flame front we need to write down an
equation for the
mean of $<dh/dt>$ given an arbitrary number $N$ of poles in the system.
This equation follows directly from (\ref{Eqdim}):
\begin{equation}
\left<{dh\over dt}\right>={1  \over L^2}{1 \over
2\pi}\int_{0}^{2\pi}u^2d\theta  \ .
\label{eqr0}
\end{equation}
After substitution of (\ref{upoles}) in (\ref{eqr0}) we get, using (\ref{xj})
and (\ref{yj})
\begin{equation}
\left<{dh\over dt}\right>=2\nu\sum_{k=1}^N {dy_k\over dt}+2\left( {\nu N\over
     L}-{\nu^2 N^2\over L^2}\right)  \ . \label{r0pole}
\end{equation}
The estimates of the second and third terms in this equation are
straightforward.
Writing $N=N(L)+\delta N(L)$ and remembering that $N(L) \sim L/\nu $
and $\delta N(L) \sim N(L)/2$
we find that
these terms contribute $O(1)$. The first term
will contribute
only when the current of poles is asymmetric. Since noise introduces poles
at a finite
value of $y_{min}$, whereas the rejected poles stream towards infinity
and disappear at boundary of nonlinearity defined by position of highest pole

\begin{equation}
y_{max}\approx 2\ln{\Big[{4L\over \pi^2\nu}\Big]} \ . \label{delN}
\end{equation}

, we have
an asymmetry that contributes to the velocity of the front. To estimate the first
term let us define

\begin{equation}
d(\sum {dy_k\over dt})=\sum_{l}^{l+dl} {dy_k\over dt}
\end{equation}

where $\sum_{l}^{l+dl} {dy_k\over dt}$ is sum over poles that are on the
interval $y:[l,l+dl]$.
We can write

\begin{equation}
d(\sum {dy_k\over dt})=d(\sum {dy_k\over dt})_{up}-
d(\sum {dy_k\over dt})_{down}
\end{equation}

Where $d(\sum {dy_k\over dt})_{up}$ flux of poles moving up and
$d(\sum {dy_k\over dt})_{down}$ flux of poles moving down.

For these flux we can write

\begin{equation}
d(\sum {dy_k\over dt})_{up},d(\sum {dy_k\over dt})_{down} \leq {dN \over
dt} dl
\end{equation}

So for the first term

\begin{eqnarray}
&&0 \leq \sum_{k=1}^{N} {dy_k\over dt}= \\ \nonumber &&
\int_{y_{min}}^{y_{max}}{d(\sum {dy_k\over dt})
\over dl} dl\\ \nonumber &&
=\int_{y_{min}}^{y_{max}}{d(\sum {dy_k\over dt})_{up}-
d(\sum {dy_k\over dt})_{down} \over dl} dl \\ \nonumber &&
\leq {dN \over dt}(y_{max}-
y_{min})\\ \nonumber &&
\leq {dN \over dt}y_{max}
\end{eqnarray}

 Because of slow($\ln$) dependence of $y_{max}$ on $L$ and $\nu$
 ${dN \over dt}$ term define oder of nonlinearity for first term.
 This term zero for symmetric current of poles and achieves maximum
 for maximal asymmetric current of poles. Comparison $v \sim L^{0.42}f^{0.02}$
 and ${dN \over dt} \sim L^{0.8}f^{0.03}$ confirm this calculation.

\section{summary and conclusions}\label{sec:method2}

The main two messages of this paper like the previous one are: (i)
There is an important interaction between the instability of
developing fronts and random noise; (ii) This interaction and its
implications can be understood qualitatively and sometimes
quantitatively using the description in terms of complex poles.

The pole description is natural in this context firstly because it
provides an
exact (and effective) representation of the steady state without noise.
Once one succeeds
to describe also the {\em perturbations} about this steady state in terms
of poles,
one achieves a particularly transparent language for the study of the
interplay between
noise and instability. This language also allows us to describe in
qualitative and
semi-quantitative terms the inverse cascade process of increasing typical
lengths when
the system relaxes to the steady state from small, random initial conditions.

The main conceptual steps in this paper are as follows: firstly one
realizes that
the steady state solution, which is characterized by $N(L)$ poles aligned along
the imaginary axis is marginally stable against noise in a periodic array of
$L$ values. For all values of $L$ the steady state is nonlinearly unstable
against noise. The main and foremost effect of noise of a given amplitude $f$ is
to introduce
an excess number of poles $\delta N(L,f)$ into the system. The existence of this
excess number of poles is responsible for the additional wrinkling of the
flame front on top of the giant cusp, and for the observed acceleration of
the flame front.
By considering the noisy appearance of new poles we rationalize the
observed scaling laws as a function of the noise amplitude and the system size.

The ``phase diagram" as a function of $L$ and $f$ in this system
consists of four regimes (in contradiction with our previous
results \cite{prev}). In the first one, discussed in Section
\ref{regime0} , the noise is too small to have any effect on the
giant cusp solution. The second regime (very small excess number
of poles ) can not be observed because of numerical noise and
discussed only theoretically. In the third regime the noise
introduces excess poles that serve to decorate the giant cusp with
side cusps. In this regime we find scaling laws for the velocity
as a function of $L$ and $f$ and we are reasonably successful in
understanding the scaling exponents. In the fourth regime the
noise is large enough to create small scale structures that are
not neatly understood in terms of individual poles. It appears
from our numerics that in this regime the roughening of the flame
front gains a contribution from the the small scale structure in a
way that is reminiscent of {\em stable}, noise driven growth
models like the Kardar-Parisi-Zhang model.

One of our main motivations in this research was to understand the
phenomena observed in radial geometry with expanding flame fronts.
A full analysis of this problem cannot be presented here. We note
however that many of the insights offered above translate
immediately to that problem. Indeed, in radial geometry the flame
front accelerates and cusps multiply and form a hierarchic
structure as time progresses. Since the radius (and the typical
scale) increase in this system all the time, new poles will be
added to the system even by a vanishingly small noise. The
marginal stability found above holds also in this case, and the
system will allow the introduction of excess poles as a result of
noise. The results discussed in Ref.\cite{96KOP} can be combined
with the present insights to provide a theory of radial growth.
This theory was offered in Ref.\cite{KOP951}.

We have had a serious open problem for this case \cite{Kupopen},
but this problem was solved successfully recently Karlin and
Sivashinsky \cite{Kar1} , \cite{Kar2}. For a cylindrical case of
the flame front propagation problem at absence of noise (only
numerical noise Ref.\cite{Seva1,Seva2,Seva3,Seva4,Seva5}) by
Sivashinsky with help of numerical methods it was shown, that the
flame front is continuously accelerated. During all this account
time it is not visible any attributes of saturation. To increase
time of the account is a difficult task. Hence, absence or
presence of velocity saturation in a cylindrical case, as
consequence of the flame front motion equation it was a open
problem before appearance of the papers \cite{Kar1} , \cite{Kar2}.

\begin{figure}
\epsfxsize=9.0truecm \epsfbox{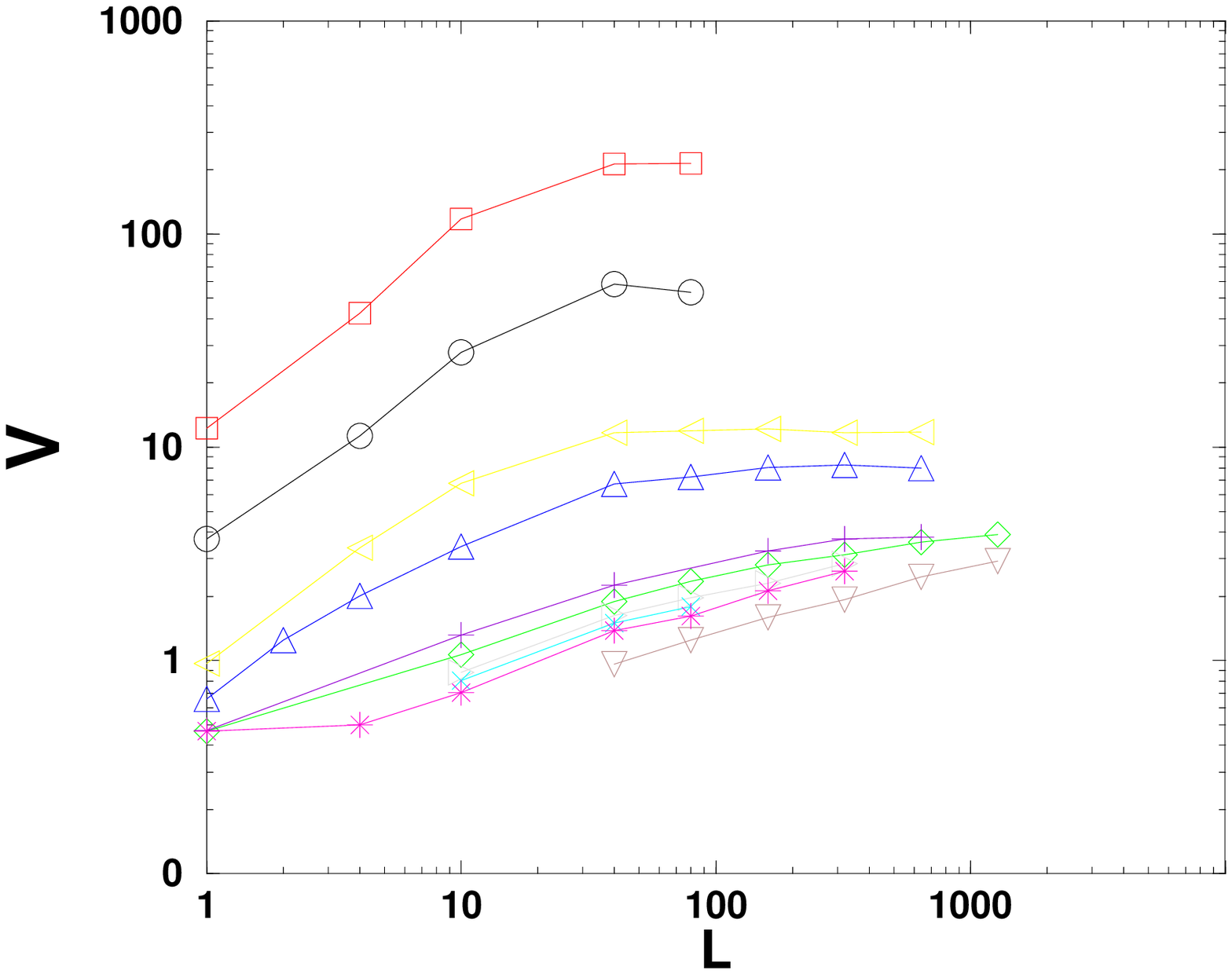} \caption {}
\label{file=Fig.1}
\end{figure}

\begin{figure}
\epsfxsize=9.0truecm \epsfbox{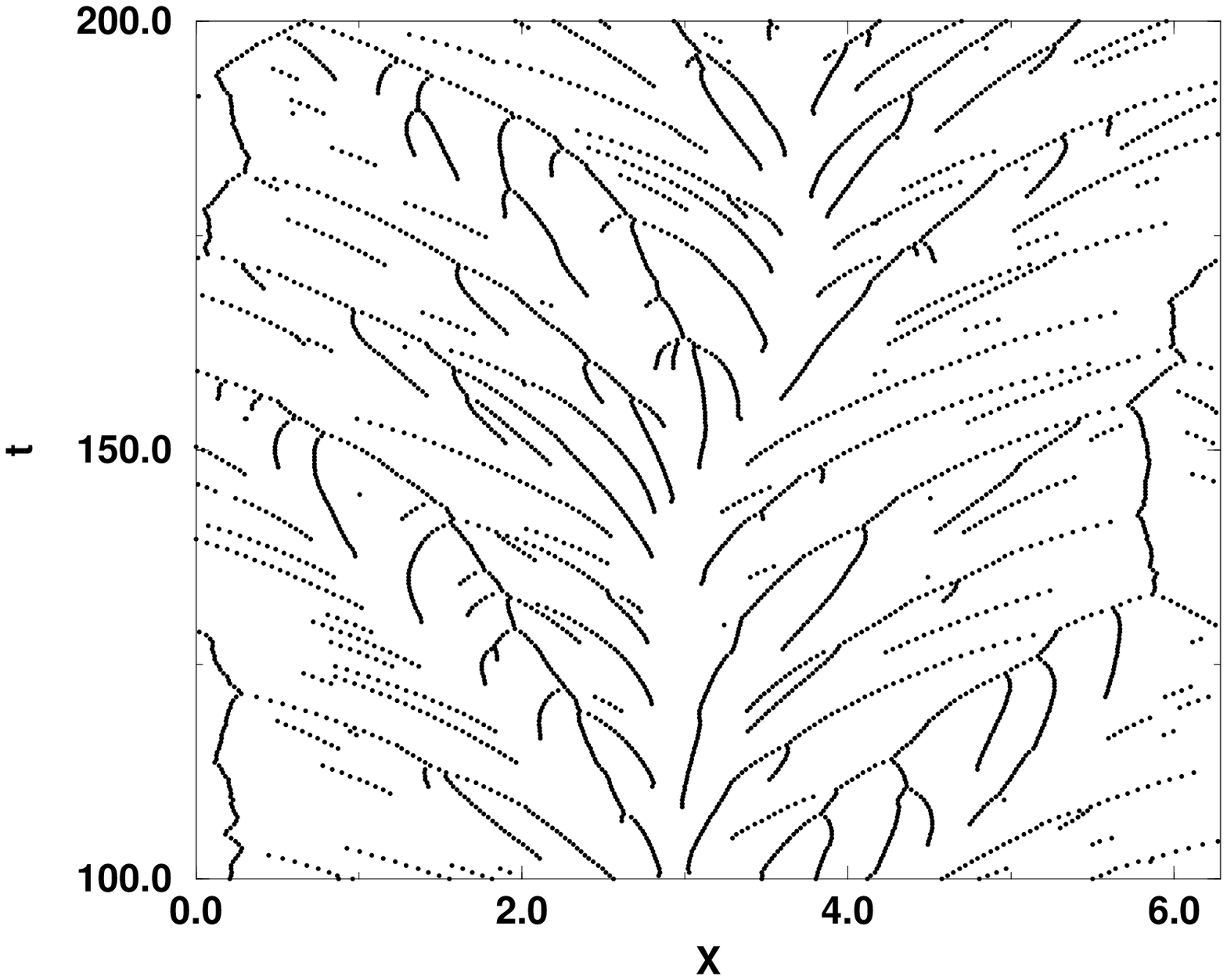} \caption {}
\label{file=Fig.3}
\end{figure}

\begin{figure}
\epsfxsize=9.0truecm \epsfbox{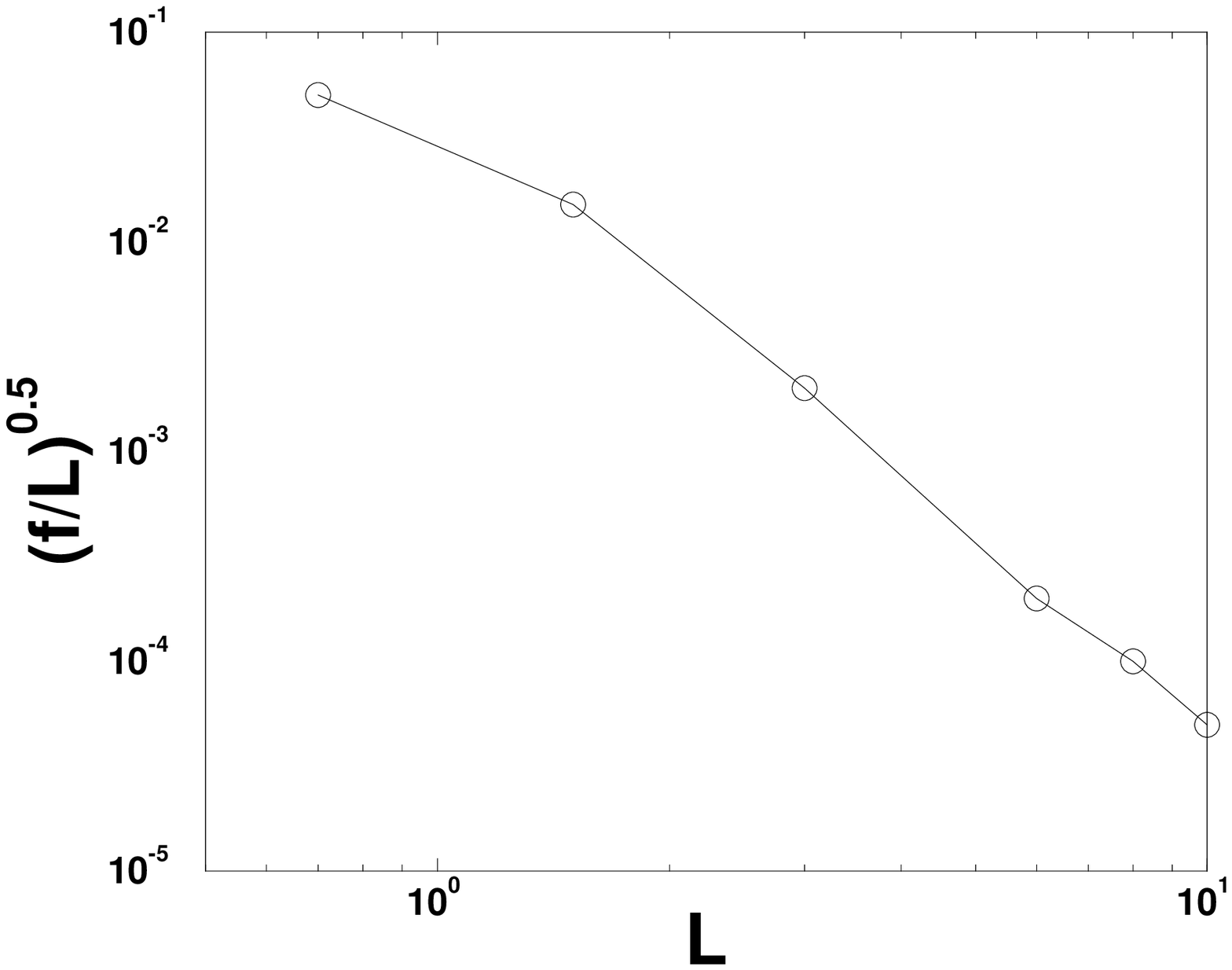} \caption {}
\label{file=Fig.4}
\end{figure}

For the best understanding of dependence of flame front velocity
as functions of its radius in a cylindrical case similar
dependence of flame front velocity on width of the channel (in a
flat case) also was analyzed by numerical methods. Growth of
velocity is also observed and at absence of noise (only numerical
noise!) also any saturation of the velocity it is not observed.
Introduction obvious Gaussian noise results to appearance of a
point of saturation and its removal from the origin of coordinates
with decreasing of noise amplitude, allowing extrapolating results
on small numerical noise. (Fig.\ref{file=Fig.1})

Hence, introducing of Gaussian noise in numerical calculation also
for a cylindrical case can again results to appearance of a
saturation point and will allow to investigate its behavior as
function of noise amplitude by extrapolating results on small
numerical noise. This investigation was really made and saturation
was observed recently by Karlin and Sivashinsky \cite{Kar1} ,
\cite{Kar2} for 1D, 2D and 3D cases.

Finally, the success of this approach in the case of flame
propagation demonstrates that Laplacian growth patterns can be
dealt with using similar ideas. A problem of immediate interest is
Laplacian growth in channels, in which a finger steady-state
solution is known to exist. It is documented that the stability of
such a finger solution to noise decreases rapidly with increasing
the channel width. In addition, it is understood that noise brings
about additional geometric features on top of the finger. There
are enough similarities here to indicate that a careful analysis
of the analytic theory may shed as much light on that problem as
on the present one.

\noindent {\bf Acknowledgments} The authors are grateful to the
anonymous referee for various suggestions to improve the clarity
of the paper. We would like also to thank Itamar Procaccia for his
supervision and many fruitful ideas result in creating the paper.
We also would like to thank Barak Galanti for
all necessary numerical calculations executed for this paper.\\~\\

\newpage

{\Large \bf Figures Legends}
\\~\\
Fig.1: The dependence of the average velocity $v$ on the noise
$f^{0.5}$ for $L$=10, 40, 80.
\\~\\
Fig.2: The dependence of the average velocity $v$ on the system
size $L$ for $f^{0.5}=0, 2.7*10^{-6}, 2.7*10^{-5}, 2.7*10^{-4},
2.7*10^{-3}, 2.7*10^{-2}, 2.7*10^{-1}$, 0.5, 1.3, 2.7 .
\\~\\
Fig.3: The dependence of the cusps positions  on time.$L=80$
$\nu=0.1$ $f=9*10{-6}$
\\~\\
Fig.4: The first odd eigenfunction obtained from traditional
stability analysis.

\newpage

\newpage

\newpage

\newpage

\end{document}